\documentclass[aps,twocolumn,showpacs,floats,longbibliography]{revtex4}
\usepackage{graphicx}
\usepackage{dcolumn}
\usepackage{bm}
\usepackage{color}
\usepackage{amsmath}
\usepackage{amssymb}
\usepackage{hyperref}

\begin{document}

\title{Localization of interacting Fermi gases in quasiperiodic potentials}

\author{Sebastiano Pilati$^{1,2}$}
\author{Vipin Kerala Varma$^{1}$}
\affiliation{$^{1}$The Abdus Salam International Centre for Theoretical Physics, 34151 Trieste, Italy}
\affiliation{$^{2}$Scuola Normale Superiore, 56126 Pisa, Italy}

\begin{abstract}
We investigate the zero-temperature metal-insulator transition in a one-dimensional two-component Fermi gas in the presence of a quasi-periodic potential resulting from the superposition of two optical 
lattices of equal intensity but incommensurate periods. 
A mobility edge separating (low energy) Anderson localized and (high energy) extended single-particle states appears in this continuous-space model beyond a critical intensity of the quasi-periodic potential.
In order to discern the metallic phase from the insulating phase in the interacting many-fermion system, we employ unbiased quantum Monte Carlo (QMC) simulations combined with the many-particle 
localization length familiar from the modern theory of the insulating state. 
In the noninteracting limit, the critical optical-lattice intensity for the metal-insulator transition predicted by the QMC simulations coincides with the Anderson localization transition of the 
single-particle eigenstates. We show that weak repulsive interactions induce a shift of this critical point towards larger intensities, meaning that repulsion favors metallic behavior. 
This shift appears to be linear in the interaction parameter, suggesting that even infinitesimal interactions can affect the position of the critical point.
\end{abstract}

\maketitle


 To what extent, if at all, do Anderson insulators persist in the presence of interactions? This has been an outstanding problem since 1958 when noninteracting quantum systems were theoretically shown 
 by Anderson to harbor no transport of conserved quantities for sufficiently strong disorder \cite{anderson1958absence, Gang4}.
In cold atom settings, among others, experimenters have observed the Anderson transition of noninteracting particles either in the nondeterministic random disorder created using spatially-correlated speckle patterns or in the 
one-dimensional quasidisorder created using incommensurate bichromatic lattice~\cite{billy2008direct,roati2008anderson,jendrzejewski2012three,kondov2011three,mcgehee2013three}. Theoretical predictions about the critical point of the Anderson transition based on models that take into account the details of these cold-atoms experiments have been recently 
reported~\cite{delande2014mobility,fratini1,fratini2,pasek2016anderson}, enabling quantitative comparison with experimental measurements~\cite{semeghini}.\\
%
%
%
%
\begin{figure}
\begin{center}
\includegraphics[width=1.0\columnwidth]{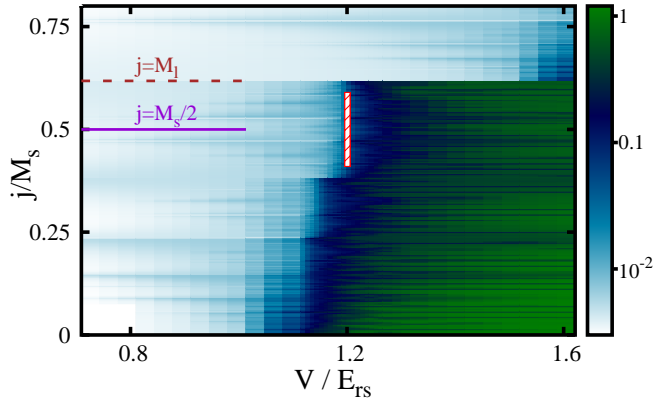}
\caption{(Color online) Logarithmic color-scale plot of the inverse participation ratio of the single-particle eigenstates $d_s/PR$ as a function of the rescaled eigenstate index $j/M_s$ and of the 
quasidisorder intensity $V/E_{rs}$, i.e. the intensity of the two optical lattices. $M_s=L/d_s=610$ and $M_l=M_s/r=377$ are the number of periods of the short-period and of the long-period lattices, respectively; 
$E_{rs}$ is the recoil energy corresponding to the former. The ratio of the two optical-lattice periods is $d_l/d_s=r\cong1.61803$ i.e. close to the golden ratio.  
The continuous horizontal (violet) segment indicates the index of the highest-occupied orbital for a density so that the short-period lattice is half filled, while the dashed (brown) horizontal segment 
the one so that the long-period lattice is fully filled. The vertical (red) bar with diagonal pattern indicates the Anderson localization transition of the states with index $j\simeq M_s/2$.}
\label{fig1}
\end{center}
\end{figure}
%
Cold-atoms experiments have emerged as the ideal playground to explore also the effects due to interactions in disordered many-body systems~\cite{aspect2009anderson,sanchez2010disordered}. 
 Experiments to understand the transport and localization phenomena in disordered interacting atomic gases continue to be performed~\cite{deissler2010delocalization,clement2006experimental,tanzi2013transport,esslinger,d2014observation,kondov2015disorder,schreiber2015observation,Choi1547}.
 Theoretically, a decade ago Basko and collaborators showed using diagrammatic techniques that the Anderson insulator can survive in the presence of interactions~\cite{Basko}, 
 even at finite temperatures. 
For continuous-space disordered bosons this finite-temperature localization~\cite{Michal} connects, in the zero temperature limit, to the superfluid to Bose glass transition~\cite{Schulz}. 
The concomitant zero-temperature localization transition for continuous-space weakly interacting quasidisordered fermions is the subject of our study.

In this Rapid Communication, we investigate the zero-temperature metal-insulator transition in a one-dimensional two-component Fermi gas with contact repulsive interactions. We consider a realistic continuous-space model for a cold-atom setup where an atomic Fermi gas is subjected to the quasiperiodic potential created by the superposition of two periodic optical lattices with the same intensity but with incommensurate 
periods. Similarly to the related (discrete-lattice) Aubry-Andr\'e model~\cite{AubryAndre} $-$ which would describe this physical system if one of the two optical lattices was very deep and the 
other extremely weak $-$ the single-particle spectrum of this (continuous-space) model displays an Anderson transition where (part of) the eigenstates become spatially localized; 
however, in contrast to the Aubry-Andr\'e model, here there is a mobility edge which separates the localized state with energies below the mobility edge, 
from the extended ergodic states above it~\cite{boers2007mobility,biddle2009localization}.

In order to discern the metallic phase from the insulating phase we adopt the concepts familiar from the modern theory of the insulating state~\cite{Kohn}, in particular the expectation value of the 
many-body phase operator~\cite{RS}. 
This approach allows one to distinguish metals from insulators by inspecting ground-state properties i.e. without direct computation of low-lying exited states or dynamical properties. 
In the interacting case, we compute this quantity via unbiased quantum Monte Carlo simulations based on the diffusion Monte Carlo algorithm, which is suitable for simulating large-scale realistic models, 
paving the way to quantitative comparison with experiments in interacting atomic gases. 
Our main goal is to inspect the effects of weak interactions on the critical point of the Anderson transition i.e. whether it drifts towards stronger or weaker intensities of the 
quasiperiodic potential, or instead if it is insensitive to interactions below a certain threshold.

The one-dimensional atomic Fermi gas we consider is described by the  following continuous-space Hamiltonian:
\begin{equation}
\hat{H} = 
       \sum_{i=1             }^{N} \left( -\frac{\hbar^2}{2m}\frac{\mathrm{d}^2}{\mathrm{d}x_{i}^2}  + v(x_i)   \right)
       +  \sum_{i_\uparrow,i_\downarrow}g\delta(x_{i_\uparrow} -x_{i_\downarrow}) 
       \;,
\label{hamiltonian}
\end{equation}
where $\hbar$ is the reduced Planck constant, $m$ is the atomic mass, $N_\uparrow$ and $N_\downarrow$ are the numbers of atoms of the two components $-$ 
hereafter referred to as spin-up and spin-down particles $-$ which are labelled by the indices $i_\uparrow = 1,\dots,N_\uparrow$ and $i_\downarrow = N_\uparrow+1,\dots,N$, respectively, 
and $N=N_\uparrow+ N_\downarrow$ is the total atom number. 
The one-dimensional coupling constant $g=-2\hbar^2/(ma_{1D})$ is related to the one-dimensional scattering length $a_{1D}$. We consider repulsive interactions $g\geqslant 0$. 
In experiments realized in tightly confining cigar-shaped waveguides, sufficiently strong to enter the regime where the gas is kinematically one-dimensional, the coupling constant $g$ can be related to 
the experimental parameters~\cite{olshanii1998atomic}, such as the three-dimensional s-wave scattering length (tunable using Feshbach resonances) and the radial harmonic confining frequency.
%
%
It is convenient to introduce the interaction parameter 
$\gamma=mg/(\hbar^2n)=2/(n|a_{1D}|)$, where $n=N/L$ is the density. The external potential $v(x)=V\left[ \sin^2\left(\pi x/d_s\right) + \sin^2\left(\pi x /d_l\right)\right]$ is the superposition of two optical lattices, 
one with the (short) period $d_s$, the other with the (long) period $d_l$. 
%
%
\begin{figure}
\begin{center}
\includegraphics[width=1.0\columnwidth]{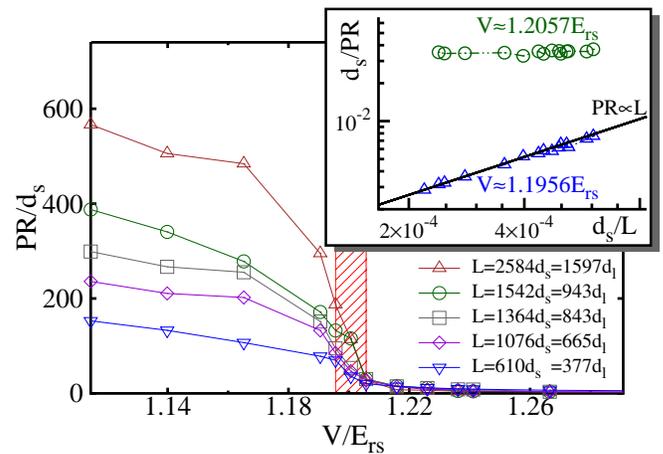}
\caption{(Color online)
Main panel: Participation ratio $PR$ of the single-particle eigenstate labelled $j=M_s/2$ as a function of the quasi-disorder strength $V/E_{rs}$, for different system sizes $L$. $d_s$ and $d_s$ are the 
period lengths of the short-period and of the long-period lattices, respectively. The vertical (red) bar with diagonal pattern indicates the location of the Anderson transition.
Inset: scaling of the inverse participation ratio $d_s/\textrm{PR}$ as a function of the inverse system size $d_s/L$, for two values of the quasidisorder strength: in one case PR 
saturates for large system sizes, whereas in the other case it diverges as $\textrm{PR}\propto L$ (see continuous black line). 
These two values bracket the critical point, and they determine the width of the (red) bar in the main panel. 
}
\label{fig2}
\end{center}
\end{figure}
\begin{figure}
\begin{center}
\includegraphics[width=1.0\columnwidth]{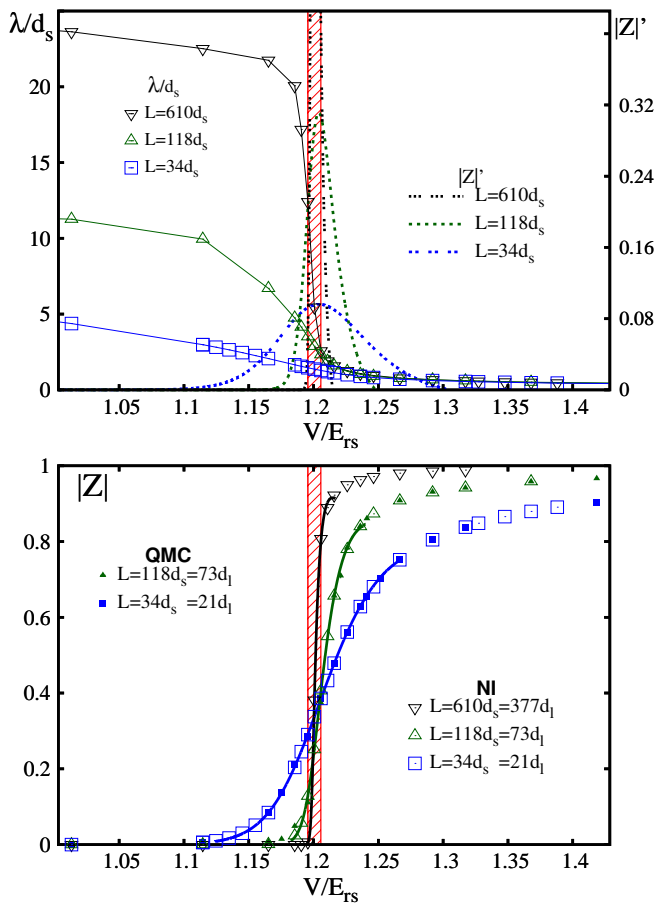}
\caption{(Color online)
Lower panel: modulus of the expectation value of the many-body phase operator $|Z|$ as a function of the quasidisorder strength $V/E_{rs}$, for different system sizes $L$. 
Full and empty symbols correspond to QMC and numerical-integration (NI) data, respectively. The continuous curves are the empirical fitting functions (see text). 
The vertical (red) bar indicates the Anderson transition. The density $n= 1/d_s$ is fixed so that the short-period lattice is half-filled, the period-lengths ratio is $r\simeq 1.618$.
Upper panel: empty-symbols with connecting lines indicate the many-particle localization length $\lambda/d_s$ (left vertical axis), while the dashed curves indicate the 
(rescaled) derivative of $|Z|$ with respect to $V/E_{rs}$ (right vertical axis).
}
\label{fig3}
\end{center}
\end{figure}

In order to form an infinite quasi-periodic potential, that is, a deterministic but aperiodic modulation, one should set the ratio of the two periods to be an irrational (Diophantine) number~\cite{modugno2009exponential}. 
However, in a finite-size continuous system such choice is incompatible with the use of periodic boundary conditions, which are in fact adopted in our calculations. 
%
The best remedy consists in choosing ratios of pairs of coprime integer numbers which, in the thermodynamic limit, converge to an irrational number. One convenient choice~\cite{diener2001transition,modugno2009exponential} is to 
set $r=d_l/d_s=K_{k+1}/K_{k}$, where the integer sequence $\{K_k\}$ (with $k=0,1,\dots$) is the Fibonacci sequence (defined by the rule $K_{k+2}=K_{k+1}+K_{k}$, with $K_0=1$ and $K_1=1$), in which 
case the limiting value for $k\rightarrow \infty$ is the golden ratio: $r\rightarrow \phi\cong1.61803$; if the system size is fixed as $L=K_{k+1}d_s=K_{k}d_l$, as we do in our 
calculations, the potential $v(x)$ complies with periodic boundary conditions, still being aperiodic within the finite box of length $L$. The intensity of the two optical lattices $V$ plays the role of 
quasi-disorder strength.
Notice that also other coprime ratios $K_{k+1}/K_{k}$, not taken from the standard Fibonacci sequence, can give similar values of period ratio 
$r\simeq \phi$, and will be considered in our analysis.

Before addressing the (interacting) many-fermion system, we inspect the properties of the single-particle eigenstates $\psi_j(x)$ of the quasi-periodic potential $v(x)$ 
(which we label with the index $j=1,2,\dots$ for increasing eigenenergies). 
We compute them by performing exact diagonalization of the finite Hamiltonian matrix obtained by introducing a 
fine discretization in the continuous-space, and approximating the second derivative in $\hat{H}$ using a finite difference formula~\cite{notepoints}.  In order to quantify the spatial extent of the single-particle eigenstates, 
we compute the normalized participation ratio $PR= \left(\int_0^L \mathrm{d}x|\psi_i(x)|^2\right)^2/\int_0^L \mathrm{d}x|\psi_i(x)|^4$. 
Ergodic extended states are characterized by large values of the participation ratio, diverging in the thermodynamic limit as $PR\propto L$ (in one-dimension); 
instead, for localized states, $PR$ is essentially independent on $L$, for sufficiently large systems~\cite{Kramer1993}. 
In Fig.~\ref{fig1} we display the $PR$ value as a function of the eigenstate index $j$ and of the disorder strength $V$. 
A sudden drop is noticeable around $V\approx 1.2 E_{rs}$ (where $E_{rs}= \pi^2\hbar^2/(2md_s^2)$ is the recoil energy of the short-period lattice, which is chosen to be the energy unit, 
while $d_s$ is used as the length unit), slightly depending on $j$, signaling an Anderson localization transition where the single-particle eigenstates become spatially localized. 
Furthermore, for $V\gtrsim1.2 E_{rs}$, a sudden increase of $PR$ for $j>M_l$ ($M_l=L/d_l$ in the number of periods of the long-period optical lattice; similarly $M_s=L/d_s$) is clearly visible, indicating a mobility edge separating the localized states with $j\leqslant M_l$, from extended states with $j>M_l$. This feature distinguishes the continuous-space model we consider from the related Aubry-Andre\'e model (i.e., a tight-binding discrete-lattice model with an incommensurate potential), where there are no mobility edges, meaning that the whole spectrum localizes at the critical quasi-disorder strength~\cite{boers2007mobility,biddle2009localization}. In fact, it has previously been found that extended Aubry-Andre\'e models which include beyond-nearest neighbor hopping processes, 
as well as other continuous-space quasi-periodic models similar to ours, host mobility edges~\cite{diener2001transition,diener2001transition,Biddle}.

Below we will consider a spin-population balanced (i.e. with $N_\uparrow=N_\downarrow=N/2$) many-fermion system with density $n=1/d_s$, meaning that the short-period lattice is half filled ($N=M_s$). At this density, the highest occupied orbital $-$ whose energy corresponds to the Fermi energy $-$ has the index $j=M_s/2$.
In order to precisely pinpoint the critical quasi-disorder strength where the Anderson localization occurs at this energy, we perform a finite-size scaling analysis of the $PR$ values; see Fig.~\ref{fig2}. In the inset, the scaling behaviors for two values of the quasi-disorder strength are shown. In order to reduce fluctuations due to finite-size effects, we average $PR$ values for $M_s/40$ states with index around $j=M_s/2$. The scaling behaviors are opposite, saturating to a finite value for the larger $V$, diverging with system size for the smaller $V$. This allows us to locate the critical point $V_c^0$ in the narrow interval between the two $V$ values: $1.1956E_{rs}<V_c^0< 1.2057E_{rs}$. According to the theory of Anderson insulators, at this critical point a metal-insulator transition occurs~\cite{Kramer1993}. 
It is worth emphasizing that the specific choice for the value of $r\simeq \phi$ is not crucial; an Anderson localization transition would occur also for different values (avoiding simple rational numbers and keeping $M_s$ large 
~\cite{AubryAndre, modugno2009exponential}), albeit at a different quasi-disorder strength~\cite{biddle2009localization}.

While the single-particle analysis reported above is suitable to identify the insulator transition in noninteracting disordered systems, for the interacting case we need a different approach. We tackle this problem by adopting the tools from the modern theory of the insulating state~\cite{Resta}. This theory was initiated by Kohn with a seminal article published in 1963~\cite{Kohn}, where he first proposed that insulating behavior in many-electron systems results from the organization of the electrons in the many-particle ground state and that insulators can be identified without inspecting exited state properties (as in the conventional theory of band insulator), nor the spatial extent of the single-particle eigenstate at the Fermi energy (as in the theory of noninteracting Anderson insulators). Resta and Sorella\cite{RS}, and later on Souza, Wilkens and Martin~\cite{SWM}, developed a rigorous formalism, which has already proven successful to identify band, Mott~\cite{RS}, as well as  Anderson insulators~\cite{RestaAnderson}, both in the case of uncorrelated random disorder and also in systems with correlated disorder, with quasi-periodic potentials, and in quasicrystals~\cite{VarmaI,VarmaII}. 
Furthermore, this formalism is amenable to powerful ab-initio computational techniques such as quantum Monte Carlo simulations~\cite{Stella,hine2007localization}.

It has emerged that in order to discern insulators from metals one has to compute the expectation value $Z = \left< \Psi \left| \hat{U} \right| \Psi \right>$ ($\left|\Psi \right>$ is the 
many-body ground state) of the many-body phase operator $\hat{U}= \exp (i (2\pi/L) \hat{X} )$, where $\hat{X}= \sum_{i=1}^N x_i $. $|Z|$ is the figure of merit to distinguish the two phases.
The theory predicts that $|Z|\rightarrow 0$ in the thermodynamic limit ($L\rightarrow \infty$ at fixed $n$) for metals, while $|Z|\rightarrow 1$ for insulators. 
Furthermore, one can define a many-particle localization length $\lambda$, which is related to the fluctuations of the macroscopic polarization, as $\lambda^2 = -\frac{L^2}{4\pi^2N}\log\left(\left|Z\right|^2\right)$. For metals, $\lambda$ diverges in the thermodynamic limit, while it saturates to a finite value for insulators, for sufficiently large systems.

In a noninteracting many-fermion system, the ground-state many-body wave-function is the Slater determinant $D(N)$ of the first $N$ single-particle spin-orbitals, which involve the first $N/2$ single-particle spatial wave-functions $\psi_i(x)$. In this case, the expectation value $Z$ is readily evaluated as $Z=\left(\det S\right)^2$ where $S$ is the $N/2\times N/2$ matrix of the overlaps $S_{ij} = \int \mathrm{d}x \psi_j^*(x)\psi_i(x)\exp\left(i 2\pi x/L\right)$. 
Alternatively, $Z$ can be computed via a QMC simulation that samples the modulus squared of the exact wave-function $\Psi_{\mathrm{NI}} (X)=D(N_\uparrow)D(N_\downarrow)$ [$X=(x_1,\dots,x_N)$ 
is the spatial configuration], where the Slater determinants of the spin-up and spin-down components are separately written for computational efficiency.
In Fig.~\ref{fig3} we show data for $|Z|$ and $\lambda$ obtained with both techniques (which we refer to as numerical integration and QMC simulation, respectively) as a function of the quasi-disorder strength $V$, for different system sizes $L$. These results confirm the expectations, in particular, $|Z|$ decreases with $L$ for small $V$, while it increases saturating to $|Z|=1$ for strong quasi-disorder. In order to pinpoint the metal-insulator transition using the finite $L$ data, where the $|Z|$ vs. $V$ curve is smooth $-$ as opposed to the thermodynamic limit, in which case a sudden jump develops $-$ 
we consider two criteria. 
The first consists in identifying the critical point with the location of the crossing of dataset corresponding to different system sizes. 
The second consists in identifying the critical point with the location of the maximum of the derivative of the curve $|Z|(V)$ with respect to $V$ (indicated as $|Z|'$ ) which, for sufficiently large $L$, would accurately approximate the position where the derivative diverges in the thermodynamic limit. 
In order to locate this point, we fit the data with an empirical fitting function based on a modified hyperbolic tangent function: 
$|Z|(V) = \frac{ \exp\left(a_1(V-c)\right) - \exp\left(-b_1(V-c)\right) }{ \exp\left(a_2(V-c) \right) - \exp\left(-b_2(V-c)\right)}$, where the $a's$, $b's$, and $c$ are fitting parameters, 
and we compute its derivative analytically. As is evident from Fig.~\ref{fig3}, both criteria provide accurate estimates of the critical quasi-disorder strength, in excellent agreement with the 
predictions based on the system-size scaling of the $PR$ values discussed above, even for the relatively small system sizes amenable to the QMC simulations.

 %
\begin{figure}
\begin{center}
\includegraphics[width=1.0\columnwidth]{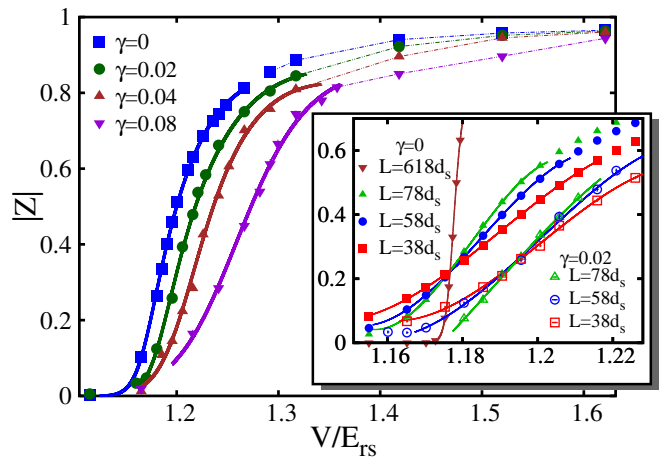}
\caption{(Color online)
Main panel: modulus of the expectation value of the many-body phase operator $|Z|$ as a function of the quasidisorder strength $V/E_{rs}$, for different interaction strengths $\gamma$. 
The density $n= 1/d_s$ is fixed so that the short-period lattice is half-filled, the period-lengths ratio is $r\simeq 1.65$; full line shows modified hyperbolic tangent (see text).
Inset: finite-size scaling analysis of $|Z|$ for the noninteracting case $\gamma=0$ (full symbols) and for an interacting case with $\gamma=0.02$ (empty symbols). Continuous curves are cubic 
fitting functions shown as guide to eye.
}
\label{fig4}
\end{center}
\end{figure}
%
\begin{figure}
\begin{center}
\includegraphics[width=1.0\columnwidth]{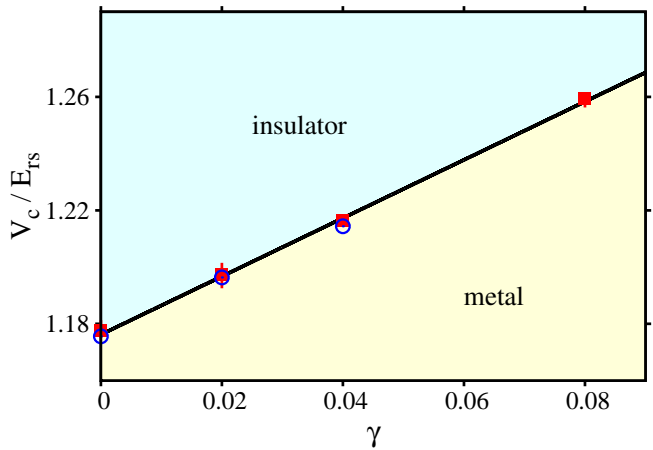}
\caption{(Color online) Critical quasidisorder strength $V_c/E_{rs}$, which separates the metallic phase (yellow region) from the insulating phase (cyan region), as a function of the interaction strength $\gamma$. 
The density is $n=1/d_s$, the optical lattice period is $r\simeq 1.65$. The empty (blue) circles indicate the data obtained from the crossing of the $|Z|(V)$  curves for different system sizes; 
the full (red) squares those obtained from the maximum of the derivative of these curves. The continuous (black) line $V_c/E_{rs}=1.03(3)\gamma+1.176(2)$ is a linear fit to the latter dataset.
}
\label{fig5}
\end{center}
\end{figure}

%
In order to determine $|Z|$ for the interacting many-fermion system, we employ QMC simulations based on the diffusion Monte Carlo algorithm~\cite{reynolds1982fixed}. 
This projective techniques stochastically solves the imaginary-time Schr\"odinger equation, and allows one to sample the exact ground-state wave function. In order to circumvent the sign-problem, 
which would hinder many-fermion simulations, one has to introduce the fixed-node constraint, meaning that the ground-state wave function is forced to have the same nodes as 
those of a trial wave function. While in generic higher-dimensional systems this constraint would possibly introduce an uncontrolled approximation, in the one-dimensional case considered here this is not 
the case, since the wave function $\Psi_{\mathrm{NI}} (X)$ defined above has the same nodes as the exact ground state~\cite{matveeva2016one}. 
Furthermore, in order to compute the unbiased expectation value of $\hat{U}$ we employ the standard forward walking technique. 
Therefore, the data reported in this Rapid Communication are free of systematic approximations. In order to reduce the stochastic fluctuations, we employ the importance sampling technique with the 
trial wave function written in the Jastrow-Slater form: $\Psi_T(X) = \Psi_{\mathrm{NI}} (X) \prod_{i_\uparrow i_\downarrow} f(\left|x_{i_\uparrow}-x_{i_\downarrow}\right|)$ which, 
beyond the Slater-determinant part $\Psi_{\mathrm{NI}} (X)$ that fixes the nodes, includes a Jastrow correlation function $f(x)>0$ that has to ensure the Bethe-Peierls boundary 
condition ${\partial \Psi}/{\partial \left(x_{i_\uparrow}-x_{i_\downarrow}\right)}\left.\right|_0=-\Psi/a_{1D}$, but is arbitrary otherwise (the specific choice affects only the statistical fluctuations).\\
In the diffusion Monte Carlo simulations, for numerical convenience we consider the system sizes $L=38d_s=23d_l$, $L=58d_s=35d_l$, and for the weakest interactions also $L=78d_s=47d_l$. $r=d_l/d_s$ is again the ratio of two coprime 
integers, but with the value $r\cong 1.65$, which is slightly larger than the golden ratio $r\cong \phi$ considered above. In the noninteracting case (for which we use the numerical integration approach) 
we also consider the size $L=618d_s=373d_l$.  Larger systems cannot be addressed via QMC simulations with the available computational resources due to the glassy nature of the insulating phase, 
which causes a pathological slow-down of the QMC dynamics and, therefore, a dramatic increase of the computational times.
In Fig.~\ref{fig4} we show the results for $|Z|$ as a function of $V$, obtained for different values of the interaction strength $\gamma$. 
As $\gamma$ increases, the datasets are shifted (approximately homogeneously for the weakest interactions) towards significantly stronger quasi-disorder; 
this clearly indicates that interactions favors metallic behavior. In order to quantify this effect, we determine the critical quasi-disorder strength $V_c$ which separates the metal from the insulator 
using the two criteria (crossings and peaks of derivative) described above in the noninteracting case. The inset of Fig.~\ref{fig5} displays the finite-size scaling analysis for the interaction parameter 
$\gamma=0.02$, compared to the noninteracting case $\gamma=0$. In the interacting case, the crossing of the curves $-$ which we identify with the critical quasi-disorder, according to the 
first criterion $-$ is clearly drifted towards larger values of $V$ compared to the noninteracting case (notice that for $r\simeq 1.65$ the metal-insulator transition occurs at slightly 
weaker quasi-disorder than in the case $r\simeq \phi$ considered before), confirming that even interactions as weak as $\gamma=0.02$ determine a positive shift of $V_c$. This is the main result of this work.\\
The zero-temperature phase diagram as a function of quasi-disorder strength $V$ and interaction parameters $\gamma$, including the metallic and the insulating phases, is displayed in Fig.~\ref{fig5}; the critical quasi-disorder strengths 
determined using the two criteria are compared, finding precise agreement. These data turn out to be well described by a simple linear fitting function. This suggests that even infinitesimal interactions 
affect the location of the metal-insulator transition. A similar linear increase of the critical quasi-disorder for weak repulsion was previously obtained for the Aubry-Andr\'e model within the 
self-consistent Hartree-Fock approximation~\cite{VarmaI}; also the statistical dynamical mean-field theory of Ref.~\cite{SemmlerHofstetter} predicts delocalizing effects due to repulsive interaction. It is 
likely that this linear increase would cease to hold for strong interactions $\gamma\gtrsim 1$; this regime is however beyond the scope of this work.

In conclusion, we have investigated the effect of weak repulsive contact interaction on the Anderson localization transition in a one-dimensional atomic Fermi gas exposed to a quasiperiodic potential. 
Our results clearly indicate that even weak repulsions induce a (seemingly linear) drift of the metal-insulator transition towards stronger quasidisorder. These results have been obtained by employing (unbiased) 
QMC simulations to compute the expectation value of the many-body phase operator familiar from the modern theory of the insulating state; this provides us with a novel approach to investigate the 
conduction properties of (quasi)disordered many-fermion systems which is suitable to address significantly larger system sizes and more complex models compared to the exact diagonalization calculations commonly adopted in the literature.
This study parallels previous investigations on ultracold atoms in shallow optical lattices~\cite{pilati2011bosonic,de2012phase,gordillo2015bosonic,astrakharchik2016one,boeris2016mott,PhysRevA.91.043618,pilati2014}, which explored
regimes where simple single-band tight binding approximations are not applicable and intriguing multi-band effects come into play.
%
In extended tight-binding models~\cite{VarmaI, VarmaIII}, interactions induce interesting effects like shifts of the critical point or, 
at finite energy-density, many-body mobility edges and nonergodic extended 
phases~\cite{Iyer, SarmaGaneshan}. The approach we implemented is a promising tool to further explore these and other phenomena, 
especially in the relatively unexplored finite temperature continuous-space setting and in higher-dimensional systems. 

We thank G. E. Astrakharchik, R. Fazio, M. Holzmann, V. E. Kravtsov, and U. Schneider for useful discussions. S. P. acknowledges financial support from the EU-H2020 project No. 641122 QUIC - Quantum simulations of 
insulators and conductors.


\begin{thebibliography}{56}%
\makeatletter
\providecommand \@ifxundefined [1]{%
 \@ifx{#1\undefined}
}%
\providecommand \@ifnum [1]{%
 \ifnum #1\expandafter \@firstoftwo
 \else \expandafter \@secondoftwo
 \fi
}%
\providecommand \@ifx [1]{%
 \ifx #1\expandafter \@firstoftwo
 \else \expandafter \@secondoftwo
 \fi
}%
\providecommand \natexlab [1]{#1}%
\providecommand \enquote  [1]{``#1''}%
\providecommand \bibnamefont  [1]{#1}%
\providecommand \bibfnamefont [1]{#1}%
\providecommand \citenamefont [1]{#1}%
\providecommand \href@noop [0]{\@secondoftwo}%
\providecommand \href [0]{\begingroup \@sanitize@url \@href}%
\providecommand \@href[1]{\@@startlink{#1}\@@href}%
\providecommand \@@href[1]{\endgroup#1\@@endlink}%
\providecommand \@sanitize@url [0]{\catcode `\\12\catcode `\$12\catcode
  `\&12\catcode `\#12\catcode `\^12\catcode `\_12\catcode `\%12\relax}%
\providecommand \@@startlink[1]{}%
\providecommand \@@endlink[0]{}%
\providecommand \url  [0]{\begingroup\@sanitize@url \@url }%
\providecommand \@url [1]{\endgroup\@href {#1}{\urlprefix }}%
\providecommand \urlprefix  [0]{URL }%
\providecommand \Eprint [0]{\href }%
\providecommand \doibase [0]{http://dx.doi.org/}%
\providecommand \selectlanguage [0]{\@gobble}%
\providecommand \bibinfo  [0]{\@secondoftwo}%
\providecommand \bibfield  [0]{\@secondoftwo}%
\providecommand \translation [1]{[#1]}%
\providecommand \BibitemOpen [0]{}%
\providecommand \bibitemStop [0]{}%
\providecommand \bibitemNoStop [0]{.\EOS\space}%
\providecommand \EOS [0]{\spacefactor3000\relax}%
\providecommand \BibitemShut  [1]{\csname bibitem#1\endcsname}%
\let\auto@bib@innerbib\@empty
\bibitem [{\citenamefont {Anderson}(1958)}]{anderson1958absence}%
  \BibitemOpen
  \bibfield  {author} {\bibinfo {author} {\bibfnamefont {P.~W.}\ \bibnamefont
  {Anderson}},\ }\href@noop {} {\bibfield  {journal} {\bibinfo  {journal}
  {Phys. Rev.}\ }\textbf {\bibinfo {volume} {109}},\ \bibinfo {pages} {1492}
  (\bibinfo {year} {1958})}\BibitemShut {NoStop}%
\bibitem [{\citenamefont {Abrahams}\ \emph {et~al.}(1979)\citenamefont
  {Abrahams}, \citenamefont {Anderson}, \citenamefont {Licciardello},\ and\
  \citenamefont {Ramakrishnan}}]{Gang4}%
  \BibitemOpen
  \bibfield  {author} {\bibinfo {author} {\bibfnamefont {E.}~\bibnamefont
  {Abrahams}}, \bibinfo {author} {\bibfnamefont {P.~W.}\ \bibnamefont
  {Anderson}}, \bibinfo {author} {\bibfnamefont {D.~C.}\ \bibnamefont
  {Licciardello}}, \ and\ \bibinfo {author} {\bibfnamefont {T.~V.}\
  \bibnamefont {Ramakrishnan}},\ }\href@noop {} {\bibfield  {journal} {\bibinfo
   {journal} {Phys. Rev. Lett.}\ }\textbf {\bibinfo {volume} {42}},\ \bibinfo
  {pages} {673} (\bibinfo {year} {1979})}\BibitemShut {NoStop}%
\bibitem [{\citenamefont {Billy}\ \emph {et~al.}(2008)\citenamefont {Billy},
  \citenamefont {Josse}, \citenamefont {Zuo}, \citenamefont {Bernard},
  \citenamefont {Hambrecht}, \citenamefont {Lugan}, \citenamefont
  {Cl{\'e}ment}, \citenamefont {Sanchez-Palencia}, \citenamefont {Bouyer},\
  and\ \citenamefont {Aspect}}]{billy2008direct}%
  \BibitemOpen
  \bibfield  {author} {\bibinfo {author} {\bibfnamefont {J.}~\bibnamefont
  {Billy}}, \bibinfo {author} {\bibfnamefont {V.}~\bibnamefont {Josse}},
  \bibinfo {author} {\bibfnamefont {Z.}~\bibnamefont {Zuo}}, \bibinfo {author}
  {\bibfnamefont {A.}~\bibnamefont {Bernard}}, \bibinfo {author} {\bibfnamefont
  {B.}~\bibnamefont {Hambrecht}}, \bibinfo {author} {\bibfnamefont
  {P.}~\bibnamefont {Lugan}}, \bibinfo {author} {\bibfnamefont
  {D.}~\bibnamefont {Cl{\'e}ment}}, \bibinfo {author} {\bibfnamefont
  {L.}~\bibnamefont {Sanchez-Palencia}}, \bibinfo {author} {\bibfnamefont
  {P.}~\bibnamefont {Bouyer}}, \ and\ \bibinfo {author} {\bibfnamefont
  {A.}~\bibnamefont {Aspect}},\ }\href@noop {} {\bibfield  {journal} {\bibinfo
  {journal} {Nature}\ }\textbf {\bibinfo {volume} {453}},\ \bibinfo {pages}
  {891} (\bibinfo {year} {2008})}\BibitemShut {NoStop}%
\bibitem [{\citenamefont {Roati}\ \emph {et~al.}(2008)\citenamefont {Roati},
  \citenamefont {D'Errico}, \citenamefont {Fallani}, \citenamefont {Fattori},
  \citenamefont {Fort}, \citenamefont {Zaccanti}, \citenamefont {Modugno},
  \citenamefont {Modugno},\ and\ \citenamefont {Inguscio}}]{roati2008anderson}%
  \BibitemOpen
  \bibfield  {author} {\bibinfo {author} {\bibfnamefont {G.}~\bibnamefont
  {Roati}}, \bibinfo {author} {\bibfnamefont {C.}~\bibnamefont {D'Errico}},
  \bibinfo {author} {\bibfnamefont {L.}~\bibnamefont {Fallani}}, \bibinfo
  {author} {\bibfnamefont {M.}~\bibnamefont {Fattori}}, \bibinfo {author}
  {\bibfnamefont {C.}~\bibnamefont {Fort}}, \bibinfo {author} {\bibfnamefont
  {M.}~\bibnamefont {Zaccanti}}, \bibinfo {author} {\bibfnamefont
  {G.}~\bibnamefont {Modugno}}, \bibinfo {author} {\bibfnamefont
  {M.}~\bibnamefont {Modugno}}, \ and\ \bibinfo {author} {\bibfnamefont
  {M.}~\bibnamefont {Inguscio}},\ }\href@noop {} {\bibfield  {journal}
  {\bibinfo  {journal} {Nature}\ }\textbf {\bibinfo {volume} {453}},\ \bibinfo
  {pages} {895} (\bibinfo {year} {2008})}\BibitemShut {NoStop}%
\bibitem [{\citenamefont {Jendrzejewski}\ \emph {et~al.}(2012)\citenamefont
  {Jendrzejewski}, \citenamefont {Bernard}, \citenamefont {Mueller},
  \citenamefont {Cheinet}, \citenamefont {Josse}, \citenamefont {Piraud},
  \citenamefont {Pezz{\'e}}, \citenamefont {Sanchez-Palencia}, \citenamefont
  {Aspect},\ and\ \citenamefont {Bouyer}}]{jendrzejewski2012three}%
  \BibitemOpen
  \bibfield  {author} {\bibinfo {author} {\bibfnamefont {F.}~\bibnamefont
  {Jendrzejewski}}, \bibinfo {author} {\bibfnamefont {A.}~\bibnamefont
  {Bernard}}, \bibinfo {author} {\bibfnamefont {K.}~\bibnamefont {Mueller}},
  \bibinfo {author} {\bibfnamefont {P.}~\bibnamefont {Cheinet}}, \bibinfo
  {author} {\bibfnamefont {V.}~\bibnamefont {Josse}}, \bibinfo {author}
  {\bibfnamefont {M.}~\bibnamefont {Piraud}}, \bibinfo {author} {\bibfnamefont
  {L.}~\bibnamefont {Pezz{\'e}}}, \bibinfo {author} {\bibfnamefont
  {L.}~\bibnamefont {Sanchez-Palencia}}, \bibinfo {author} {\bibfnamefont
  {A.}~\bibnamefont {Aspect}}, \ and\ \bibinfo {author} {\bibfnamefont
  {P.}~\bibnamefont {Bouyer}},\ }\href@noop {} {\bibfield  {journal} {\bibinfo
  {journal} {Nat. Phys.}\ }\textbf {\bibinfo {volume} {8}},\ \bibinfo {pages}
  {398} (\bibinfo {year} {2012})}\BibitemShut {NoStop}%
\bibitem [{\citenamefont {Kondov}\ \emph {et~al.}(2011)\citenamefont {Kondov},
  \citenamefont {McGehee}, \citenamefont {Zirbel},\ and\ \citenamefont
  {DeMarco}}]{kondov2011three}%
  \BibitemOpen
  \bibfield  {author} {\bibinfo {author} {\bibfnamefont {S.}~\bibnamefont
  {Kondov}}, \bibinfo {author} {\bibfnamefont {W.}~\bibnamefont {McGehee}},
  \bibinfo {author} {\bibfnamefont {J.}~\bibnamefont {Zirbel}}, \ and\ \bibinfo
  {author} {\bibfnamefont {B.}~\bibnamefont {DeMarco}},\ }\href@noop {}
  {\bibfield  {journal} {\bibinfo  {journal} {Science}\ }\textbf {\bibinfo
  {volume} {334}},\ \bibinfo {pages} {66} (\bibinfo {year} {2011})}\BibitemShut
  {NoStop}%
\bibitem [{\citenamefont {McGehee}\ \emph {et~al.}(2013)\citenamefont
  {McGehee}, \citenamefont {Kondov}, \citenamefont {Xu}, \citenamefont
  {Zirbel},\ and\ \citenamefont {DeMarco}}]{mcgehee2013three}%
  \BibitemOpen
  \bibfield  {author} {\bibinfo {author} {\bibfnamefont {W.}~\bibnamefont
  {McGehee}}, \bibinfo {author} {\bibfnamefont {S.}~\bibnamefont {Kondov}},
  \bibinfo {author} {\bibfnamefont {W.}~\bibnamefont {Xu}}, \bibinfo {author}
  {\bibfnamefont {J.}~\bibnamefont {Zirbel}}, \ and\ \bibinfo {author}
  {\bibfnamefont {B.}~\bibnamefont {DeMarco}},\ }\href@noop {} {\bibfield
  {journal} {\bibinfo  {journal} {Phys. Rev. Lett.}\ }\textbf {\bibinfo
  {volume} {111}},\ \bibinfo {pages} {145303} (\bibinfo {year}
  {2013})}\BibitemShut {NoStop}%
\bibitem [{\citenamefont {Delande}\ and\ \citenamefont
  {Orso}(2014)}]{delande2014mobility}%
  \BibitemOpen
  \bibfield  {author} {\bibinfo {author} {\bibfnamefont {D.}~\bibnamefont
  {Delande}}\ and\ \bibinfo {author} {\bibfnamefont {G.}~\bibnamefont {Orso}},\
  }\href@noop {} {\bibfield  {journal} {\bibinfo  {journal} {Phys. Rev. Lett.}\
  }\textbf {\bibinfo {volume} {113}},\ \bibinfo {pages} {060601} (\bibinfo
  {year} {2014})}\BibitemShut {NoStop}%
\bibitem [{\citenamefont {Fratini}\ and\ \citenamefont
  {Pilati}(2015{\natexlab{a}})}]{fratini1}%
  \BibitemOpen
  \bibfield  {author} {\bibinfo {author} {\bibfnamefont {E.}~\bibnamefont
  {Fratini}}\ and\ \bibinfo {author} {\bibfnamefont {S.}~\bibnamefont
  {Pilati}},\ }\href@noop {} {\bibfield  {journal} {\bibinfo  {journal} {Phys.
  Rev. A}\ }\textbf {\bibinfo {volume} {91}},\ \bibinfo {pages} {061601}
  (\bibinfo {year} {2015}{\natexlab{a}})}\BibitemShut {NoStop}%
\bibitem [{\citenamefont {Fratini}\ and\ \citenamefont
  {Pilati}(2015{\natexlab{b}})}]{fratini2}%
  \BibitemOpen
  \bibfield  {author} {\bibinfo {author} {\bibfnamefont {E.}~\bibnamefont
  {Fratini}}\ and\ \bibinfo {author} {\bibfnamefont {S.}~\bibnamefont
  {Pilati}},\ }\href@noop {} {\bibfield  {journal} {\bibinfo  {journal} {Phys.
  Rev. A}\ }\textbf {\bibinfo {volume} {92}},\ \bibinfo {pages} {063621}
  (\bibinfo {year} {2015}{\natexlab{b}})}\BibitemShut {NoStop}%
\bibitem [{\citenamefont {Pasek}\ \emph {et~al.}(2016)\citenamefont {Pasek},
  \citenamefont {Orso},\ and\ \citenamefont {Delande}}]{pasek2016anderson}%
  \BibitemOpen
  \bibfield  {author} {\bibinfo {author} {\bibfnamefont {M.}~\bibnamefont
  {Pasek}}, \bibinfo {author} {\bibfnamefont {G.}~\bibnamefont {Orso}}, \ and\
  \bibinfo {author} {\bibfnamefont {D.}~\bibnamefont {Delande}},\ }\href@noop
  {} {\bibfield  {journal} {\bibinfo  {journal} {arXiv preprint
  arXiv:1609.01065}\ } (\bibinfo {year} {2016})}\BibitemShut {NoStop}%
\bibitem [{\citenamefont {Semeghini}\ \emph {et~al.}(2015)\citenamefont
  {Semeghini}, \citenamefont {Landini}, \citenamefont {Castilho}, \citenamefont
  {Roy}, \citenamefont {Spagnolli}, \citenamefont {Trenkwalder}, \citenamefont
  {Fattori}, \citenamefont {Inguscio},\ and\ \citenamefont
  {Modugno}}]{semeghini}%
  \BibitemOpen
  \bibfield  {author} {\bibinfo {author} {\bibfnamefont {G.}~\bibnamefont
  {Semeghini}}, \bibinfo {author} {\bibfnamefont {M.}~\bibnamefont {Landini}},
  \bibinfo {author} {\bibfnamefont {P.}~\bibnamefont {Castilho}}, \bibinfo
  {author} {\bibfnamefont {S.}~\bibnamefont {Roy}}, \bibinfo {author}
  {\bibfnamefont {G.}~\bibnamefont {Spagnolli}}, \bibinfo {author}
  {\bibfnamefont {A.}~\bibnamefont {Trenkwalder}}, \bibinfo {author}
  {\bibfnamefont {M.}~\bibnamefont {Fattori}}, \bibinfo {author} {\bibfnamefont
  {M.}~\bibnamefont {Inguscio}}, \ and\ \bibinfo {author} {\bibfnamefont
  {G.}~\bibnamefont {Modugno}},\ }\href@noop {} {\bibfield  {journal} {\bibinfo
   {journal} {Nat. Phys.}\ }\textbf {\bibinfo {volume} {11}},\ \bibinfo {pages}
  {554} (\bibinfo {year} {2015})}\BibitemShut {NoStop}%
\bibitem [{\citenamefont {Aspect}\ and\ \citenamefont
  {Inguscio}(2009)}]{aspect2009anderson}%
  \BibitemOpen
  \bibfield  {author} {\bibinfo {author} {\bibfnamefont {A.}~\bibnamefont
  {Aspect}}\ and\ \bibinfo {author} {\bibfnamefont {M.}~\bibnamefont
  {Inguscio}},\ }\href@noop {} {\bibfield  {journal} {\bibinfo  {journal}
  {Phys. Today}\ }\textbf {\bibinfo {volume} {62}},\ \bibinfo {pages} {30}
  (\bibinfo {year} {2009})}\BibitemShut {NoStop}%
\bibitem [{\citenamefont {Sanchez-Palencia}\ and\ \citenamefont
  {Lewenstein}(2010)}]{sanchez2010disordered}%
  \BibitemOpen
  \bibfield  {author} {\bibinfo {author} {\bibfnamefont {L.}~\bibnamefont
  {Sanchez-Palencia}}\ and\ \bibinfo {author} {\bibfnamefont {M.}~\bibnamefont
  {Lewenstein}},\ }\href@noop {} {\bibfield  {journal} {\bibinfo  {journal}
  {Nat. Phys.}\ }\textbf {\bibinfo {volume} {6}},\ \bibinfo {pages} {87}
  (\bibinfo {year} {2010})}\BibitemShut {NoStop}%
\bibitem [{\citenamefont {Deissler}\ \emph {et~al.}(2010)\citenamefont
  {Deissler}, \citenamefont {Zaccanti}, \citenamefont {Roati}, \citenamefont
  {D'Errico}, \citenamefont {Fattori}, \citenamefont {Modugno}, \citenamefont
  {Modugno},\ and\ \citenamefont {Inguscio}}]{deissler2010delocalization}%
  \BibitemOpen
  \bibfield  {author} {\bibinfo {author} {\bibfnamefont {B.}~\bibnamefont
  {Deissler}}, \bibinfo {author} {\bibfnamefont {M.}~\bibnamefont {Zaccanti}},
  \bibinfo {author} {\bibfnamefont {G.}~\bibnamefont {Roati}}, \bibinfo
  {author} {\bibfnamefont {C.}~\bibnamefont {D'Errico}}, \bibinfo {author}
  {\bibfnamefont {M.}~\bibnamefont {Fattori}}, \bibinfo {author} {\bibfnamefont
  {M.}~\bibnamefont {Modugno}}, \bibinfo {author} {\bibfnamefont
  {G.}~\bibnamefont {Modugno}}, \ and\ \bibinfo {author} {\bibfnamefont
  {M.}~\bibnamefont {Inguscio}},\ }\href@noop {} {\bibfield  {journal}
  {\bibinfo  {journal} {Nat. Phys.}\ }\textbf {\bibinfo {volume} {6}},\
  \bibinfo {pages} {354} (\bibinfo {year} {2010})}\BibitemShut {NoStop}%
\bibitem [{\citenamefont {Cl{\'e}ment}\ \emph {et~al.}(2006)\citenamefont
  {Cl{\'e}ment}, \citenamefont {Var{\`o}n}, \citenamefont {Retter},
  \citenamefont {Sanchez-Palencia}, \citenamefont {Aspect},\ and\ \citenamefont
  {Bouyer}}]{clement2006experimental}%
  \BibitemOpen
  \bibfield  {author} {\bibinfo {author} {\bibfnamefont {D.}~\bibnamefont
  {Cl{\'e}ment}}, \bibinfo {author} {\bibfnamefont {A.~F.}\ \bibnamefont
  {Var{\`o}n}}, \bibinfo {author} {\bibfnamefont {J.~A.}\ \bibnamefont
  {Retter}}, \bibinfo {author} {\bibfnamefont {L.}~\bibnamefont
  {Sanchez-Palencia}}, \bibinfo {author} {\bibfnamefont {A.}~\bibnamefont
  {Aspect}}, \ and\ \bibinfo {author} {\bibfnamefont {P.}~\bibnamefont
  {Bouyer}},\ }\href@noop {} {\bibfield  {journal} {\bibinfo  {journal} {New J.
  Phys.}\ }\textbf {\bibinfo {volume} {8}},\ \bibinfo {pages} {165} (\bibinfo
  {year} {2006})}\BibitemShut {NoStop}%
\bibitem [{\citenamefont {Tanzi}\ \emph {et~al.}(2013)\citenamefont {Tanzi},
  \citenamefont {Lucioni}, \citenamefont {Chaudhuri}, \citenamefont {Gori},
  \citenamefont {Kumar}, \citenamefont {D'Errico}, \citenamefont {Inguscio},\
  and\ \citenamefont {Modugno}}]{tanzi2013transport}%
  \BibitemOpen
  \bibfield  {author} {\bibinfo {author} {\bibfnamefont {L.}~\bibnamefont
  {Tanzi}}, \bibinfo {author} {\bibfnamefont {E.}~\bibnamefont {Lucioni}},
  \bibinfo {author} {\bibfnamefont {S.}~\bibnamefont {Chaudhuri}}, \bibinfo
  {author} {\bibfnamefont {L.}~\bibnamefont {Gori}}, \bibinfo {author}
  {\bibfnamefont {A.}~\bibnamefont {Kumar}}, \bibinfo {author} {\bibfnamefont
  {C.}~\bibnamefont {D'Errico}}, \bibinfo {author} {\bibfnamefont
  {M.}~\bibnamefont {Inguscio}}, \ and\ \bibinfo {author} {\bibfnamefont
  {G.}~\bibnamefont {Modugno}},\ }\href@noop {} {\bibfield  {journal} {\bibinfo
   {journal} {Phys. Rev. Lett.}\ }\textbf {\bibinfo {volume} {111}},\ \bibinfo
  {pages} {115301} (\bibinfo {year} {2013})}\BibitemShut {NoStop}%
\bibitem [{\citenamefont {Krinner}\ \emph {et~al.}(2015)\citenamefont
  {Krinner}, \citenamefont {Stadler}, \citenamefont {Meineke}, \citenamefont
  {Brantut},\ and\ \citenamefont {Esslinger}}]{esslinger}%
  \BibitemOpen
  \bibfield  {author} {\bibinfo {author} {\bibfnamefont {S.}~\bibnamefont
  {Krinner}}, \bibinfo {author} {\bibfnamefont {D.}~\bibnamefont {Stadler}},
  \bibinfo {author} {\bibfnamefont {J.}~\bibnamefont {Meineke}}, \bibinfo
  {author} {\bibfnamefont {J.-P.}\ \bibnamefont {Brantut}}, \ and\ \bibinfo
  {author} {\bibfnamefont {T.}~\bibnamefont {Esslinger}},\ }\href@noop {}
  {\bibfield  {journal} {\bibinfo  {journal} {Phys. Rev. Lett.}\ }\textbf
  {\bibinfo {volume} {115}},\ \bibinfo {pages} {045302} (\bibinfo {year}
  {2015})}\BibitemShut {NoStop}%
\bibitem [{\citenamefont {D'Errico}\ \emph {et~al.}(2014)\citenamefont
  {D'Errico}, \citenamefont {Lucioni}, \citenamefont {Tanzi}, \citenamefont
  {Gori}, \citenamefont {Roux}, \citenamefont {McCulloch}, \citenamefont
  {Giamarchi}, \citenamefont {Inguscio},\ and\ \citenamefont
  {Modugno}}]{d2014observation}%
  \BibitemOpen
  \bibfield  {author} {\bibinfo {author} {\bibfnamefont {C.}~\bibnamefont
  {D'Errico}}, \bibinfo {author} {\bibfnamefont {E.}~\bibnamefont {Lucioni}},
  \bibinfo {author} {\bibfnamefont {L.}~\bibnamefont {Tanzi}}, \bibinfo
  {author} {\bibfnamefont {L.}~\bibnamefont {Gori}}, \bibinfo {author}
  {\bibfnamefont {G.}~\bibnamefont {Roux}}, \bibinfo {author} {\bibfnamefont
  {I.~P.}\ \bibnamefont {McCulloch}}, \bibinfo {author} {\bibfnamefont
  {T.}~\bibnamefont {Giamarchi}}, \bibinfo {author} {\bibfnamefont
  {M.}~\bibnamefont {Inguscio}}, \ and\ \bibinfo {author} {\bibfnamefont
  {G.}~\bibnamefont {Modugno}},\ }\href@noop {} {\bibfield  {journal} {\bibinfo
   {journal} {Phys. Rev. Lett.}\ }\textbf {\bibinfo {volume} {113}},\ \bibinfo
  {pages} {095301} (\bibinfo {year} {2014})}\BibitemShut {NoStop}%
\bibitem [{\citenamefont {Kondov}\ \emph {et~al.}(2015)\citenamefont {Kondov},
  \citenamefont {McGehee}, \citenamefont {Xu},\ and\ \citenamefont
  {DeMarco}}]{kondov2015disorder}%
  \BibitemOpen
  \bibfield  {author} {\bibinfo {author} {\bibfnamefont {S.}~\bibnamefont
  {Kondov}}, \bibinfo {author} {\bibfnamefont {W.}~\bibnamefont {McGehee}},
  \bibinfo {author} {\bibfnamefont {W.}~\bibnamefont {Xu}}, \ and\ \bibinfo
  {author} {\bibfnamefont {B.}~\bibnamefont {DeMarco}},\ }\href@noop {}
  {\bibfield  {journal} {\bibinfo  {journal} {Phys. Rev. Lett.}\ }\textbf
  {\bibinfo {volume} {114}},\ \bibinfo {pages} {083002} (\bibinfo {year}
  {2015})}\BibitemShut {NoStop}%
\bibitem [{\citenamefont {Schreiber}\ \emph {et~al.}(2015)\citenamefont
  {Schreiber}, \citenamefont {Hodgman}, \citenamefont {Bordia}, \citenamefont
  {L{\"u}schen}, \citenamefont {Fischer}, \citenamefont {Vosk}, \citenamefont
  {Altman}, \citenamefont {Schneider},\ and\ \citenamefont
  {Bloch}}]{schreiber2015observation}%
  \BibitemOpen
  \bibfield  {author} {\bibinfo {author} {\bibfnamefont {M.}~\bibnamefont
  {Schreiber}}, \bibinfo {author} {\bibfnamefont {S.~S.}\ \bibnamefont
  {Hodgman}}, \bibinfo {author} {\bibfnamefont {P.}~\bibnamefont {Bordia}},
  \bibinfo {author} {\bibfnamefont {H.~P.}\ \bibnamefont {L{\"u}schen}},
  \bibinfo {author} {\bibfnamefont {M.~H.}\ \bibnamefont {Fischer}}, \bibinfo
  {author} {\bibfnamefont {R.}~\bibnamefont {Vosk}}, \bibinfo {author}
  {\bibfnamefont {E.}~\bibnamefont {Altman}}, \bibinfo {author} {\bibfnamefont
  {U.}~\bibnamefont {Schneider}}, \ and\ \bibinfo {author} {\bibfnamefont
  {I.}~\bibnamefont {Bloch}},\ }\href@noop {} {\bibfield  {journal} {\bibinfo
  {journal} {Science}\ }\textbf {\bibinfo {volume} {349}},\ \bibinfo {pages}
  {842} (\bibinfo {year} {2015})}\BibitemShut {NoStop}%
\bibitem [{\citenamefont {Choi}\ \emph {et~al.}(2016)\citenamefont {Choi},
  \citenamefont {Hild}, \citenamefont {Zeiher}, \citenamefont {Schau{\ss}},
  \citenamefont {Rubio-Abadal}, \citenamefont {Yefsah}, \citenamefont
  {Khemani}, \citenamefont {Huse}, \citenamefont {Bloch},\ and\ \citenamefont
  {Gross}}]{Choi1547}%
  \BibitemOpen
  \bibfield  {author} {\bibinfo {author} {\bibfnamefont {J.-y.}\ \bibnamefont
  {Choi}}, \bibinfo {author} {\bibfnamefont {S.}~\bibnamefont {Hild}}, \bibinfo
  {author} {\bibfnamefont {J.}~\bibnamefont {Zeiher}}, \bibinfo {author}
  {\bibfnamefont {P.}~\bibnamefont {Schau{\ss}}}, \bibinfo {author}
  {\bibfnamefont {A.}~\bibnamefont {Rubio-Abadal}}, \bibinfo {author}
  {\bibfnamefont {T.}~\bibnamefont {Yefsah}}, \bibinfo {author} {\bibfnamefont
  {V.}~\bibnamefont {Khemani}}, \bibinfo {author} {\bibfnamefont {D.~A.}\
  \bibnamefont {Huse}}, \bibinfo {author} {\bibfnamefont {I.}~\bibnamefont
  {Bloch}}, \ and\ \bibinfo {author} {\bibfnamefont {C.}~\bibnamefont
  {Gross}},\ }\href {\doibase 10.1126/science.aaf8834} {\bibfield  {journal}
  {\bibinfo  {journal} {Science}\ }\textbf {\bibinfo {volume} {352}},\ \bibinfo
  {pages} {1547} (\bibinfo {year} {2016})}\BibitemShut {NoStop}%
\bibitem [{\citenamefont {Basko}\ \emph {et~al.}(2006)\citenamefont {Basko},
  \citenamefont {Aleiner},\ and\ \citenamefont {Altshuler}}]{Basko}%
  \BibitemOpen
  \bibfield  {author} {\bibinfo {author} {\bibfnamefont {D.~M.}\ \bibnamefont
  {Basko}}, \bibinfo {author} {\bibfnamefont {I.~L.}\ \bibnamefont {Aleiner}},
  \ and\ \bibinfo {author} {\bibfnamefont {B.~L.}\ \bibnamefont {Altshuler}},\
  }\href@noop {} {\bibfield  {journal} {\bibinfo  {journal} {Ann.~Phys.
  (Amsterdam)}\ }\textbf {\bibinfo {volume} {321}},\ \bibinfo {pages} {1126}
  (\bibinfo {year} {2006})}\BibitemShut {NoStop}%
\bibitem [{\citenamefont {Michal}\ \emph {et~al.}(2016)\citenamefont {Michal},
  \citenamefont {Aleiner}, \citenamefont {Altshuler},\ and\ \citenamefont
  {Shlyapnikov}}]{Michal}%
  \BibitemOpen
  \bibfield  {author} {\bibinfo {author} {\bibfnamefont {V.~P.}\ \bibnamefont
  {Michal}}, \bibinfo {author} {\bibfnamefont {I.~L.}\ \bibnamefont {Aleiner}},
  \bibinfo {author} {\bibfnamefont {B.~L.}\ \bibnamefont {Altshuler}}, \ and\
  \bibinfo {author} {\bibfnamefont {G.~V.}\ \bibnamefont {Shlyapnikov}},\
  }\href@noop {} {\bibfield  {journal} {\bibinfo  {journal} {PNAS}\ }\textbf
  {\bibinfo {volume} {113}},\ \bibinfo {pages} {E4455} (\bibinfo {year}
  {2016})}\BibitemShut {NoStop}%
\bibitem [{\citenamefont {Giamarchi}\ and\ \citenamefont
  {Schulz}(1988)}]{Schulz}%
  \BibitemOpen
  \bibfield  {author} {\bibinfo {author} {\bibfnamefont {T.}~\bibnamefont
  {Giamarchi}}\ and\ \bibinfo {author} {\bibfnamefont {H.~J.}\ \bibnamefont
  {Schulz}},\ }\href@noop {} {\bibfield  {journal} {\bibinfo  {journal} {Phys.
  Rev. B}\ }\textbf {\bibinfo {volume} {37}},\ \bibinfo {pages} {325} (\bibinfo
  {year} {1988})}\BibitemShut {NoStop}%
\bibitem [{\citenamefont {Aubry}\ and\ \citenamefont
  {Andr{\'e}}(1980)}]{AubryAndre}%
  \BibitemOpen
  \bibfield  {author} {\bibinfo {author} {\bibfnamefont {S.}~\bibnamefont
  {Aubry}}\ and\ \bibinfo {author} {\bibfnamefont {G.}~\bibnamefont
  {Andr{\'e}}},\ }\href@noop {} {\bibfield  {journal} {\bibinfo  {journal}
  {Ann. Isr. Phys. Soc.}\ }\textbf {\bibinfo {volume} {3}},\ \bibinfo {pages}
  {133} (\bibinfo {year} {1980})}\BibitemShut {NoStop}%
\bibitem [{\citenamefont {Boers}\ \emph {et~al.}(2007)\citenamefont {Boers},
  \citenamefont {Goedeke}, \citenamefont {Hinrichs},\ and\ \citenamefont
  {Holthaus}}]{boers2007mobility}%
  \BibitemOpen
  \bibfield  {author} {\bibinfo {author} {\bibfnamefont {D.~J.}\ \bibnamefont
  {Boers}}, \bibinfo {author} {\bibfnamefont {B.}~\bibnamefont {Goedeke}},
  \bibinfo {author} {\bibfnamefont {D.}~\bibnamefont {Hinrichs}}, \ and\
  \bibinfo {author} {\bibfnamefont {M.}~\bibnamefont {Holthaus}},\ }\href@noop
  {} {\bibfield  {journal} {\bibinfo  {journal} {Phys. Rev. A}\ }\textbf
  {\bibinfo {volume} {75}},\ \bibinfo {pages} {063404} (\bibinfo {year}
  {2007})}\BibitemShut {NoStop}%
\bibitem [{\citenamefont {Biddle}\ \emph {et~al.}(2009)\citenamefont {Biddle},
  \citenamefont {Wang}, \citenamefont {Priour~Jr},\ and\ \citenamefont
  {Sarma}}]{biddle2009localization}%
  \BibitemOpen
  \bibfield  {author} {\bibinfo {author} {\bibfnamefont {J.}~\bibnamefont
  {Biddle}}, \bibinfo {author} {\bibfnamefont {B.}~\bibnamefont {Wang}},
  \bibinfo {author} {\bibfnamefont {D.}~\bibnamefont {Priour~Jr}}, \ and\
  \bibinfo {author} {\bibfnamefont {S.~D.}\ \bibnamefont {Sarma}},\ }\href@noop
  {} {\bibfield  {journal} {\bibinfo  {journal} {Phys. Rev. A}\ }\textbf
  {\bibinfo {volume} {80}},\ \bibinfo {pages} {021603} (\bibinfo {year}
  {2009})}\BibitemShut {NoStop}%
\bibitem [{\citenamefont {Kohn}(1963)}]{Kohn}%
  \BibitemOpen
  \bibfield  {author} {\bibinfo {author} {\bibfnamefont {W.}~\bibnamefont
  {Kohn}},\ }\href@noop {} {\bibfield  {journal} {\bibinfo  {journal} {Phys.
  Rev.}\ }\textbf {\bibinfo {volume} {133}},\ \bibinfo {pages} {A171} (\bibinfo
  {year} {1963})}\BibitemShut {NoStop}%
\bibitem [{\citenamefont {Resta}\ and\ \citenamefont {Sorella}(1999)}]{RS}%
  \BibitemOpen
  \bibfield  {author} {\bibinfo {author} {\bibfnamefont {R.}~\bibnamefont
  {Resta}}\ and\ \bibinfo {author} {\bibfnamefont {S.}~\bibnamefont
  {Sorella}},\ }\href@noop {} {\bibfield  {journal} {\bibinfo  {journal} {Phys.
  Rev. Lett.}\ }\textbf {\bibinfo {volume} {82}},\ \bibinfo {pages} {370}
  (\bibinfo {year} {1999})}\BibitemShut {NoStop}%
\bibitem [{\citenamefont {Olshanii}(1998)}]{olshanii1998atomic}%
  \BibitemOpen
  \bibfield  {author} {\bibinfo {author} {\bibfnamefont {M.}~\bibnamefont
  {Olshanii}},\ }\href@noop {} {\bibfield  {journal} {\bibinfo  {journal}
  {Phys. Rev. Lett.}\ }\textbf {\bibinfo {volume} {81}},\ \bibinfo {pages}
  {938} (\bibinfo {year} {1998})}\BibitemShut {NoStop}%
\bibitem [{\citenamefont {Modugno}(2009)}]{modugno2009exponential}%
  \BibitemOpen
  \bibfield  {author} {\bibinfo {author} {\bibfnamefont {M.}~\bibnamefont
  {Modugno}},\ }\href@noop {} {\bibfield  {journal} {\bibinfo  {journal} {New
  J. Phys.}\ }\textbf {\bibinfo {volume} {11}},\ \bibinfo {pages} {033023}
  (\bibinfo {year} {2009})}\BibitemShut {NoStop}%
\bibitem [{\citenamefont {Diener}\ \emph {et~al.}(2001)\citenamefont {Diener},
  \citenamefont {Georgakis}, \citenamefont {Zhong}, \citenamefont {Raizen},\
  and\ \citenamefont {Niu}}]{diener2001transition}%
  \BibitemOpen
  \bibfield  {author} {\bibinfo {author} {\bibfnamefont {R.~B.}\ \bibnamefont
  {Diener}}, \bibinfo {author} {\bibfnamefont {G.~A.}\ \bibnamefont
  {Georgakis}}, \bibinfo {author} {\bibfnamefont {J.}~\bibnamefont {Zhong}},
  \bibinfo {author} {\bibfnamefont {M.}~\bibnamefont {Raizen}}, \ and\ \bibinfo
  {author} {\bibfnamefont {Q.}~\bibnamefont {Niu}},\ }\href@noop {} {\bibfield
  {journal} {\bibinfo  {journal} {Phys. Rev. A}\ }\textbf {\bibinfo {volume}
  {64}},\ \bibinfo {pages} {033416} (\bibinfo {year} {2001})}\BibitemShut
  {NoStop}%
\bibitem [{not()}]{notepoints}%
  \BibitemOpen
  \href@noop {} {}\bibinfo {note} {We use an $11$-point formula and $20-30$
  points per $d_s$, so that the discretization error is
  negligible.}\BibitemShut {Stop}%
\bibitem [{\citenamefont {Kramer}\ and\ \citenamefont
  {MacKinnon}(1993)}]{Kramer1993}%
  \BibitemOpen
  \bibfield  {author} {\bibinfo {author} {\bibfnamefont {B.}~\bibnamefont
  {Kramer}}\ and\ \bibinfo {author} {\bibfnamefont {A.}~\bibnamefont
  {MacKinnon}},\ }\href@noop {} {\bibfield  {journal} {\bibinfo  {journal}
  {Rep. Prog. Phys.}\ }\textbf {\bibinfo {volume} {56}},\ \bibinfo {pages}
  {1469} (\bibinfo {year} {1993})}\BibitemShut {NoStop}%
\bibitem [{\citenamefont {Biddle}\ \emph {et~al.}(2011)\citenamefont {Biddle},
  \citenamefont {Priour}, \citenamefont {Wang},\ and\ \citenamefont
  {Das~Sarma}}]{Biddle}%
  \BibitemOpen
  \bibfield  {author} {\bibinfo {author} {\bibfnamefont {J.}~\bibnamefont
  {Biddle}}, \bibinfo {author} {\bibfnamefont {D.~J.}\ \bibnamefont {Priour}},
  \bibinfo {author} {\bibfnamefont {B.}~\bibnamefont {Wang}}, \ and\ \bibinfo
  {author} {\bibfnamefont {S.}~\bibnamefont {Das~Sarma}},\ }\href@noop {}
  {\bibfield  {journal} {\bibinfo  {journal} {Phys. Rev. B}\ }\textbf {\bibinfo
  {volume} {83}},\ \bibinfo {pages} {075105} (\bibinfo {year}
  {2011})}\BibitemShut {NoStop}%
\bibitem [{\citenamefont {Resta}(2011)}]{Resta}%
  \BibitemOpen
  \bibfield  {author} {\bibinfo {author} {\bibfnamefont {R.}~\bibnamefont
  {Resta}},\ }\href@noop {} {\bibfield  {journal} {\bibinfo  {journal} {Eur.
  Phys. J. B}\ }\textbf {\bibinfo {volume} {79}},\ \bibinfo {pages} {121}
  (\bibinfo {year} {2011})}\BibitemShut {NoStop}%
\bibitem [{\citenamefont {Souza}\ \emph {et~al.}(2000)\citenamefont {Souza},
  \citenamefont {Wilkens},\ and\ \citenamefont {Martin}}]{SWM}%
  \BibitemOpen
  \bibfield  {author} {\bibinfo {author} {\bibfnamefont {I.}~\bibnamefont
  {Souza}}, \bibinfo {author} {\bibfnamefont {T.}~\bibnamefont {Wilkens}}, \
  and\ \bibinfo {author} {\bibfnamefont {R.~M.}\ \bibnamefont {Martin}},\
  }\href@noop {} {\bibfield  {journal} {\bibinfo  {journal} {Phys. Rev. B}\
  }\textbf {\bibinfo {volume} {62}},\ \bibinfo {pages} {1666} (\bibinfo {year}
  {2000})}\BibitemShut {NoStop}%
\bibitem [{\citenamefont {Bendazzoli}\ \emph {et~al.}(2010)\citenamefont
  {Bendazzoli}, \citenamefont {Evangelisti}, \citenamefont {Monari},\ and\
  \citenamefont {Resta}}]{RestaAnderson}%
  \BibitemOpen
  \bibfield  {author} {\bibinfo {author} {\bibfnamefont {G.~L.}\ \bibnamefont
  {Bendazzoli}}, \bibinfo {author} {\bibfnamefont {S.}~\bibnamefont
  {Evangelisti}}, \bibinfo {author} {\bibfnamefont {A.}~\bibnamefont {Monari}},
  \ and\ \bibinfo {author} {\bibfnamefont {R.}~\bibnamefont {Resta}},\
  }\href@noop {} {\bibfield  {journal} {\bibinfo  {journal} {J. Chem. Phys.}\
  }\textbf {\bibinfo {volume} {133}},\ \bibinfo {pages} {064703} (\bibinfo
  {year} {2010})}\BibitemShut {NoStop}%
\bibitem [{\citenamefont {Kerala~Varma}\ and\ \citenamefont
  {Pilati}(2015)}]{VarmaI}%
  \BibitemOpen
  \bibfield  {author} {\bibinfo {author} {\bibfnamefont {V.}~\bibnamefont
  {Kerala~Varma}}\ and\ \bibinfo {author} {\bibfnamefont {S.}~\bibnamefont
  {Pilati}},\ }\href@noop {} {\bibfield  {journal} {\bibinfo  {journal} {Phys.
  Rev. B}\ }\textbf {\bibinfo {volume} {92}},\ \bibinfo {pages} {134207}
  (\bibinfo {year} {2015})}\BibitemShut {NoStop}%
\bibitem [{\citenamefont {Kerala~Varma}\ \emph {et~al.}(2016)\citenamefont
  {Kerala~Varma}, \citenamefont {Pilati},\ and\ \citenamefont
  {Kravtsov}}]{VarmaII}%
  \BibitemOpen
  \bibfield  {author} {\bibinfo {author} {\bibfnamefont {V.}~\bibnamefont
  {Kerala~Varma}}, \bibinfo {author} {\bibfnamefont {S.}~\bibnamefont
  {Pilati}}, \ and\ \bibinfo {author} {\bibfnamefont {V.~E.}\ \bibnamefont
  {Kravtsov}},\ }\href@noop {} {\bibfield  {journal} {\bibinfo  {journal}
  {arXiv preprint arXiv:1607.06276}\ } (\bibinfo {year} {2016})}\BibitemShut
  {NoStop}%
\bibitem [{\citenamefont {Stella}\ \emph {et~al.}(2011)\citenamefont {Stella},
  \citenamefont {Attaccalite}, \citenamefont {Sorella},\ and\ \citenamefont
  {Rubio}}]{Stella}%
  \BibitemOpen
  \bibfield  {author} {\bibinfo {author} {\bibfnamefont {L.}~\bibnamefont
  {Stella}}, \bibinfo {author} {\bibfnamefont {C.}~\bibnamefont {Attaccalite}},
  \bibinfo {author} {\bibfnamefont {S.}~\bibnamefont {Sorella}}, \ and\
  \bibinfo {author} {\bibfnamefont {A.}~\bibnamefont {Rubio}},\ }\href@noop {}
  {\bibfield  {journal} {\bibinfo  {journal} {Phys. Rev. B}\ }\textbf {\bibinfo
  {volume} {84}},\ \bibinfo {pages} {245117} (\bibinfo {year}
  {2011})}\BibitemShut {NoStop}%
\bibitem [{\citenamefont {Hine}\ and\ \citenamefont
  {Foulkes}(2007)}]{hine2007localization}%
  \BibitemOpen
  \bibfield  {author} {\bibinfo {author} {\bibfnamefont {N.}~\bibnamefont
  {Hine}}\ and\ \bibinfo {author} {\bibfnamefont {W.}~\bibnamefont {Foulkes}},\
  }\href@noop {} {\bibfield  {journal} {\bibinfo  {journal} {J. Phys.: Condens.
  Matter}\ }\textbf {\bibinfo {volume} {19}},\ \bibinfo {pages} {506212}
  (\bibinfo {year} {2007})}\BibitemShut {NoStop}%
\bibitem [{\citenamefont {Reynolds}\ \emph {et~al.}(1982)\citenamefont
  {Reynolds}, \citenamefont {Ceperley}, \citenamefont {Alder},\ and\
  \citenamefont {Lester~Jr}}]{reynolds1982fixed}%
  \BibitemOpen
  \bibfield  {author} {\bibinfo {author} {\bibfnamefont {P.~J.}\ \bibnamefont
  {Reynolds}}, \bibinfo {author} {\bibfnamefont {D.~M.}\ \bibnamefont
  {Ceperley}}, \bibinfo {author} {\bibfnamefont {B.~J.}\ \bibnamefont {Alder}},
  \ and\ \bibinfo {author} {\bibfnamefont {W.~A.}\ \bibnamefont {Lester~Jr}},\
  }\href@noop {} {\bibfield  {journal} {\bibinfo  {journal} {J. Chem. Phys.}\
  }\textbf {\bibinfo {volume} {77}},\ \bibinfo {pages} {5593} (\bibinfo {year}
  {1982})}\BibitemShut {NoStop}%
\bibitem [{\citenamefont {Matveeva}\ and\ \citenamefont
  {Astrakharchik}(2016)}]{matveeva2016one}%
  \BibitemOpen
  \bibfield  {author} {\bibinfo {author} {\bibfnamefont {N.}~\bibnamefont
  {Matveeva}}\ and\ \bibinfo {author} {\bibfnamefont {G.}~\bibnamefont
  {Astrakharchik}},\ }\href@noop {} {\bibfield  {journal} {\bibinfo  {journal}
  {arXiv preprint arXiv:1603.08794}\ } (\bibinfo {year} {2016})}\BibitemShut
  {NoStop}%
\bibitem [{\citenamefont {Semmler}\ \emph {et~al.}(2011)\citenamefont
  {Semmler}, \citenamefont {Byczuk},\ and\ \citenamefont
  {Hofstetter}}]{SemmlerHofstetter}%
  \BibitemOpen
  \bibfield  {author} {\bibinfo {author} {\bibfnamefont {D.}~\bibnamefont
  {Semmler}}, \bibinfo {author} {\bibfnamefont {K.}~\bibnamefont {Byczuk}}, \
  and\ \bibinfo {author} {\bibfnamefont {W.}~\bibnamefont {Hofstetter}},\
  }\href@noop {} {\bibfield  {journal} {\bibinfo  {journal} {Phys. Rev. B}\
  }\textbf {\bibinfo {volume} {84}},\ \bibinfo {pages} {115113} (\bibinfo
  {year} {2011})}\BibitemShut {NoStop}%
\bibitem [{\citenamefont {Pilati}\ and\ \citenamefont
  {Troyer}(2011)}]{pilati2011bosonic}%
  \BibitemOpen
  \bibfield  {author} {\bibinfo {author} {\bibfnamefont {S.}~\bibnamefont
  {Pilati}}\ and\ \bibinfo {author} {\bibfnamefont {M.}~\bibnamefont
  {Troyer}},\ }\href@noop {} {\bibfield  {journal} {\bibinfo  {journal} {Phys.
  Rev. Lett.}\ }\textbf {\bibinfo {volume} {108}},\ \bibinfo {pages} {155301}
  (\bibinfo {year} {2011})}\BibitemShut {NoStop}%
\bibitem [{\citenamefont {De~Soto}\ and\ \citenamefont
  {Gordillo}(2012)}]{de2012phase}%
  \BibitemOpen
  \bibfield  {author} {\bibinfo {author} {\bibfnamefont {F.}~\bibnamefont
  {De~Soto}}\ and\ \bibinfo {author} {\bibfnamefont {M.}~\bibnamefont
  {Gordillo}},\ }\href@noop {} {\bibfield  {journal} {\bibinfo  {journal}
  {Phys. Rev. A}\ }\textbf {\bibinfo {volume} {85}},\ \bibinfo {pages} {013607}
  (\bibinfo {year} {2012})}\BibitemShut {NoStop}%
\bibitem [{\citenamefont {Gordillo}\ \emph
  {et~al.}(2015{\natexlab{a}})\citenamefont {Gordillo}, \citenamefont
  {Carbonell-Coronado},\ and\ \citenamefont {De~Soto}}]{gordillo2015bosonic}%
  \BibitemOpen
  \bibfield  {author} {\bibinfo {author} {\bibfnamefont {M.}~\bibnamefont
  {Gordillo}}, \bibinfo {author} {\bibfnamefont {C.}~\bibnamefont
  {Carbonell-Coronado}}, \ and\ \bibinfo {author} {\bibfnamefont
  {F.}~\bibnamefont {De~Soto}},\ }\href@noop {} {\bibfield  {journal} {\bibinfo
   {journal} {Phys. Rev. A}\ }\textbf {\bibinfo {volume} {91}},\ \bibinfo
  {pages} {043618} (\bibinfo {year} {2015}{\natexlab{a}})}\BibitemShut
  {NoStop}%
\bibitem [{\citenamefont {Astrakharchik}\ \emph {et~al.}(2016)\citenamefont
  {Astrakharchik}, \citenamefont {Krutitsky}, \citenamefont {Lewenstein},\ and\
  \citenamefont {Mazzanti}}]{astrakharchik2016one}%
  \BibitemOpen
  \bibfield  {author} {\bibinfo {author} {\bibfnamefont {G.~E.}\ \bibnamefont
  {Astrakharchik}}, \bibinfo {author} {\bibfnamefont {K.~V.}\ \bibnamefont
  {Krutitsky}}, \bibinfo {author} {\bibfnamefont {M.}~\bibnamefont
  {Lewenstein}}, \ and\ \bibinfo {author} {\bibfnamefont {F.}~\bibnamefont
  {Mazzanti}},\ }\href@noop {} {\bibfield  {journal} {\bibinfo  {journal}
  {Phys. Rev. A}\ }\textbf {\bibinfo {volume} {93}},\ \bibinfo {pages} {021605}
  (\bibinfo {year} {2016})}\BibitemShut {NoStop}%
\bibitem [{\citenamefont {Bo{\'e}ris}\ \emph {et~al.}(2016)\citenamefont
  {Bo{\'e}ris}, \citenamefont {Gori}, \citenamefont {Hoogerland}, \citenamefont
  {Kumar}, \citenamefont {Lucioni}, \citenamefont {Tanzi}, \citenamefont
  {Inguscio}, \citenamefont {Giamarchi}, \citenamefont {D'Errico},
  \citenamefont {Carleo} \emph {et~al.}}]{boeris2016mott}%
  \BibitemOpen
  \bibfield  {author} {\bibinfo {author} {\bibfnamefont {G.}~\bibnamefont
  {Bo{\'e}ris}}, \bibinfo {author} {\bibfnamefont {L.}~\bibnamefont {Gori}},
  \bibinfo {author} {\bibfnamefont {M.~D.}\ \bibnamefont {Hoogerland}},
  \bibinfo {author} {\bibfnamefont {A.}~\bibnamefont {Kumar}}, \bibinfo
  {author} {\bibfnamefont {E.}~\bibnamefont {Lucioni}}, \bibinfo {author}
  {\bibfnamefont {L.}~\bibnamefont {Tanzi}}, \bibinfo {author} {\bibfnamefont
  {M.}~\bibnamefont {Inguscio}}, \bibinfo {author} {\bibfnamefont
  {T.}~\bibnamefont {Giamarchi}}, \bibinfo {author} {\bibfnamefont
  {C.}~\bibnamefont {D'Errico}}, \bibinfo {author} {\bibfnamefont
  {G.}~\bibnamefont {Carleo}},  \emph {et~al.},\ }\href@noop {} {\bibfield
  {journal} {\bibinfo  {journal} {Phys. Rev. A}\ }\textbf {\bibinfo {volume}
  {93}},\ \bibinfo {pages} {011601} (\bibinfo {year} {2016})}\BibitemShut
  {NoStop}%
\bibitem [{\citenamefont {Gordillo}\ \emph
  {et~al.}(2015{\natexlab{b}})\citenamefont {Gordillo}, \citenamefont
  {Carbonell-Coronado},\ and\ \citenamefont {De~Soto}}]{PhysRevA.91.043618}%
  \BibitemOpen
  \bibfield  {author} {\bibinfo {author} {\bibfnamefont {M.~C.}\ \bibnamefont
  {Gordillo}}, \bibinfo {author} {\bibfnamefont {C.}~\bibnamefont
  {Carbonell-Coronado}}, \ and\ \bibinfo {author} {\bibfnamefont
  {F.}~\bibnamefont {De~Soto}},\ }\href@noop {} {\bibfield  {journal} {\bibinfo
   {journal} {Phys. Rev. A}\ }\textbf {\bibinfo {volume} {91}},\ \bibinfo
  {pages} {043618} (\bibinfo {year} {2015}{\natexlab{b}})}\BibitemShut
  {NoStop}%
\bibitem [{\citenamefont {Pilati}\ \emph {et~al.}(2014)\citenamefont {Pilati},
  \citenamefont {Zintchenko},\ and\ \citenamefont {Troyer}}]{pilati2014}%
  \BibitemOpen
  \bibfield  {author} {\bibinfo {author} {\bibfnamefont {S.}~\bibnamefont
  {Pilati}}, \bibinfo {author} {\bibfnamefont {I.}~\bibnamefont {Zintchenko}},
  \ and\ \bibinfo {author} {\bibfnamefont {M.}~\bibnamefont {Troyer}},\
  }\href@noop {} {\bibfield  {journal} {\bibinfo  {journal} {Phys. Rev. Lett.}\
  }\textbf {\bibinfo {volume} {112}},\ \bibinfo {pages} {015301} (\bibinfo
  {year} {2014})}\BibitemShut {NoStop}%
\bibitem [{\citenamefont {Kerala~Varma}\ and\ \citenamefont
  {S\'anchez}(2015)}]{VarmaIII}%
  \BibitemOpen
  \bibfield  {author} {\bibinfo {author} {\bibfnamefont {V.}~\bibnamefont
  {Kerala~Varma}}\ and\ \bibinfo {author} {\bibfnamefont {R.~J.}\ \bibnamefont
  {S\'anchez}},\ }\href@noop {} {\bibfield  {journal} {\bibinfo  {journal}
  {Phys. Rev. A}\ }\textbf {\bibinfo {volume} {92}},\ \bibinfo {pages} {013618}
  (\bibinfo {year} {2015})}\BibitemShut {NoStop}%
\bibitem [{\citenamefont {Iyer}\ \emph {et~al.}(2013)\citenamefont {Iyer},
  \citenamefont {Oganesyan}, \citenamefont {Refael},\ and\ \citenamefont
  {Huse}}]{Iyer}%
  \BibitemOpen
  \bibfield  {author} {\bibinfo {author} {\bibfnamefont {S.}~\bibnamefont
  {Iyer}}, \bibinfo {author} {\bibfnamefont {V.}~\bibnamefont {Oganesyan}},
  \bibinfo {author} {\bibfnamefont {G.}~\bibnamefont {Refael}}, \ and\ \bibinfo
  {author} {\bibfnamefont {D.~A.}\ \bibnamefont {Huse}},\ }\href@noop {}
  {\bibfield  {journal} {\bibinfo  {journal} {Phys. Rev. B}\ }\textbf {\bibinfo
  {volume} {87}},\ \bibinfo {pages} {134202} (\bibinfo {year}
  {2013})}\BibitemShut {NoStop}%
\bibitem [{\citenamefont {Li}\ \emph {et~al.}(2015)\citenamefont {Li},
  \citenamefont {Ganeshan}, \citenamefont {Pixley},\ and\ \citenamefont
  {Das~Sarma}}]{SarmaGaneshan}%
  \BibitemOpen
  \bibfield  {author} {\bibinfo {author} {\bibfnamefont {X.}~\bibnamefont
  {Li}}, \bibinfo {author} {\bibfnamefont {S.}~\bibnamefont {Ganeshan}},
  \bibinfo {author} {\bibfnamefont {J.~H.}\ \bibnamefont {Pixley}}, \ and\
  \bibinfo {author} {\bibfnamefont {S.}~\bibnamefont {Das~Sarma}},\ }\href@noop
  {} {\bibfield  {journal} {\bibinfo  {journal} {Phys. Rev. Lett.}\ }\textbf
  {\bibinfo {volume} {115}},\ \bibinfo {pages} {186601} (\bibinfo {year}
  {2015})}\BibitemShut {NoStop}%
\end{thebibliography}

%
\end{document}